\documentclass[pra,aps,twocolumn,floatfix,10pt]{revtex4-1}
\usepackage{amsmath,amssymb,graphicx,amsthm,dsfont,bm,color,ulem}

\begin{document}

\title{Spontaneous localization in self-focusing of ultrashort light pulses}

\author{Miguel A. Porras}

\affiliation{Grupo de Sistemas Complejos, ETSIME, Universidad Polit\'ecnica de Madrid, Rios Rosas 21, 28003 Madrid, Spain}

\begin{abstract}
This is a summary directed to PhD students of the research work conducted on the problem of the production of ``light bullets", or multidimensional wave packets that propagate without distortion in unbounded, homogeneous, nonlinear media, and on the actual nature of the self-localized light wave packets spontaneously generated in self-focusing experiments with ultrashort pulses.
\end{abstract}

\maketitle

\section{Introduction}

This research work aims to shed some light on one of the most fundamental problems of nonlinear optics: the production of localized wave packets of light that overcome their natural tendency to spread, propagating without modification in a stable way, and thus as if they were particles of light, or ``light bullets". This problem was first posed by Chiao \textit{et al.} \cite{CHIAO} in the sixties for optical beams, and with the development of ultrashort laser sources in the nineties, renewed by Silberberg \cite{SILBERBERG1} for optical pulses.

At low intensity, spatially localized waves (beams) spread due to diffraction, and temporally localized waves (pulses) increase their duration due to the chromatic dispersion of optical materials. At high intensity, the Kerr effect results in an opposite tendency to compression that can be played against diffraction and dispersion. For waves in one dimension (a pulse in an optical fiber, for example) these opposite tendencies find a stable balance in one-dimensional (1D) solitons, which have found wide application in technology. However, multidimensional solitons propagating in an unbounded and homogeneous optical medium have been found to be much more elusive. This problem is related to the phenomenon of collapse. Contrary to the 1D case, self-focusing in 2D and 3D ends with the formation of a singularity ---a point with infinite intensity--- at a finite propagation distance. A singularity have been never seen in practical settings because material responses other than the Kerr effect turn on at the enormous intensities reached and remove the singularity. The problem of the production of a multidimensional self-localized and stable wave packet thus amounts to find suitable conditions for the effects arresting collapse to result in a stable in a non-spreading wave packet \cite{SILBERBERG2,THEHUNT,SKARKA,FIBICH,HENZ}.

This problem is closely connected to the experiments of self-channeling, self-trapping, or filamentation of light \cite{COUAIRON}. Self-focusing of pulses with a duration ranging from few hundreds of femtoseconds down to few femtoseconds, and with peak power above the critical power of self-focusing (e.g., several tens of megawatts in glass), ends with the production of a (or several) light filament accompanied by a plasma channel, that can propagate for hundred of the diffraction length that is expected from its width. The nature of these light filaments is not completely understood. The apparently self-trapped propagation conceals a very complex spatiotemporal dynamics that is hardly understood from the idea of a multidimensional soliton \cite{KOLESIK,TRAPANI3}.

These problems have led to investigate other types of multidimensional localized waves that have broadened the spectrum of possible light bullets. Non-diffracting and non-dispersing wave packets as Bessel beams and X waves \cite{DURNIN,LINEARXWAVES} and other types of conical waves \cite{PORRASEX,PORRASMODES} are linear waves that have found their counterpart in the nonlinear optics realm. A large number of nonlinear self-localized waves in two and three dimensions propagating in media with different nonlinearities have been described \cite{PORRASUBB,PORRASOWAVES1,PORRASOWAVES2,PORRASAPB}, and many of them have been found useful for understanding the behavior of light filaments, reconciling their static and dynamic aspects. \cite{TRAPANI3,PORRASY,PORRASXYZ,PORRASSP,PORRASEXX}

In this paper, I review the different types of multidimensional, self-localized waves in nonlinear media, classify them and discuss on their stability. For simplicity, details are only presented for the simplest members of the different families (those with maximum symmetry and in media with the fundamental nonlinearities). Particular emphasis is made on self-localized wave packets with the capability of propagating without any attenuation and distortion in nonlinear absorbing media, and on wave packets that self-reconstruct after obstacles.
Their spontaneous generation and their role in filamentation experiments is discussed in the last section.

\section{Subluminal solitons, superluminal conical waves, and luminal hybrids}

To simplify as far as possible, let me consider the simplest situation of a wave packet $E=A\exp[-i(\omega_0t-k_0z)]$ of the envelope $A$ propagating in a nonlinear medium according to the nonlinear Schr\"odinger
equation
\begin{equation}\label{NLSE}
\partial_z A=\frac{i}{2k_0}\Delta A  +\frac{ik_0n_2}{n_0}|A|^2A\, ,
\end{equation}
where $k_0=n_0\omega_0/c$ is the propagation constant at the carrier frequency $\omega_0$, and $n_0$ is the refractive index also at $\omega_0$. For a monochromatic light beam, the term $\Delta=\partial^2_x+\partial^2_y$ accounts for diffraction spreading in the transverse plane $(x,y)$. The cubic or Kerr term with the nonlinear refractive index $n_2$ (usually positive) describes an increase of the refraction index with intensity that causes self-compression of the wave packet.

For pulses, chromatic dispersion is accounted for up to its lowest order of approximation by setting $\Delta=\partial^2_x+\partial^2_y -k_0k_0^{\prime\prime}\partial^2_t$, where $k_0^{\prime\prime}$ is the chromatic dispersion coefficient. Also, other linear and nonlinear terms due to higher-order dispersion, saturation of Kerr nonlinearity, or dissipation of energy must be introduced in the NLSE for extremely short and intense pulses, but these do not alter substantially our discussion.

In self-localized and stable wave packets, self-compression and linear spreading (and other additional effects) find a stable balance. Mathematically, they must be found among the bounded and stationary solutions of the NLSE, or solutions $A=A(x,y)\exp(i\delta z)$  that tend to zero with increasing $(x,y)$ and that do not change with $z$ except for, possibly, a phase.
The wave vector shift $\delta$ of these solutions is usually written in the form
\begin{equation}
\delta=\alpha \frac{k_0n_2I_0}{n_0}
\end{equation}
where $I_0$ is the maximum intensity of the solution at its center. The NLSE has bounded and stationary solutions with a discrete spectrum of positive values of $\alpha$, and with a continuous spectrum of negative values of $\alpha$.

\begin{figure}
\centering
\includegraphics[width=4cm]{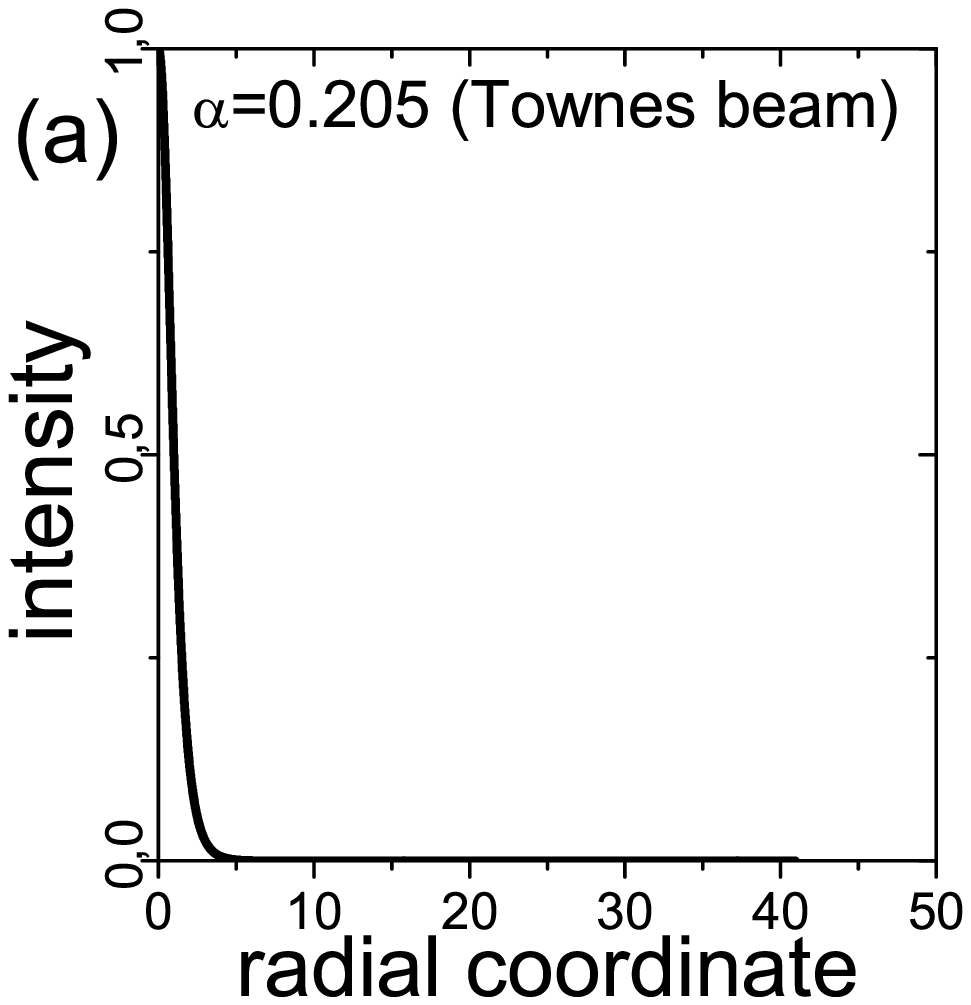}\includegraphics[width=4cm]{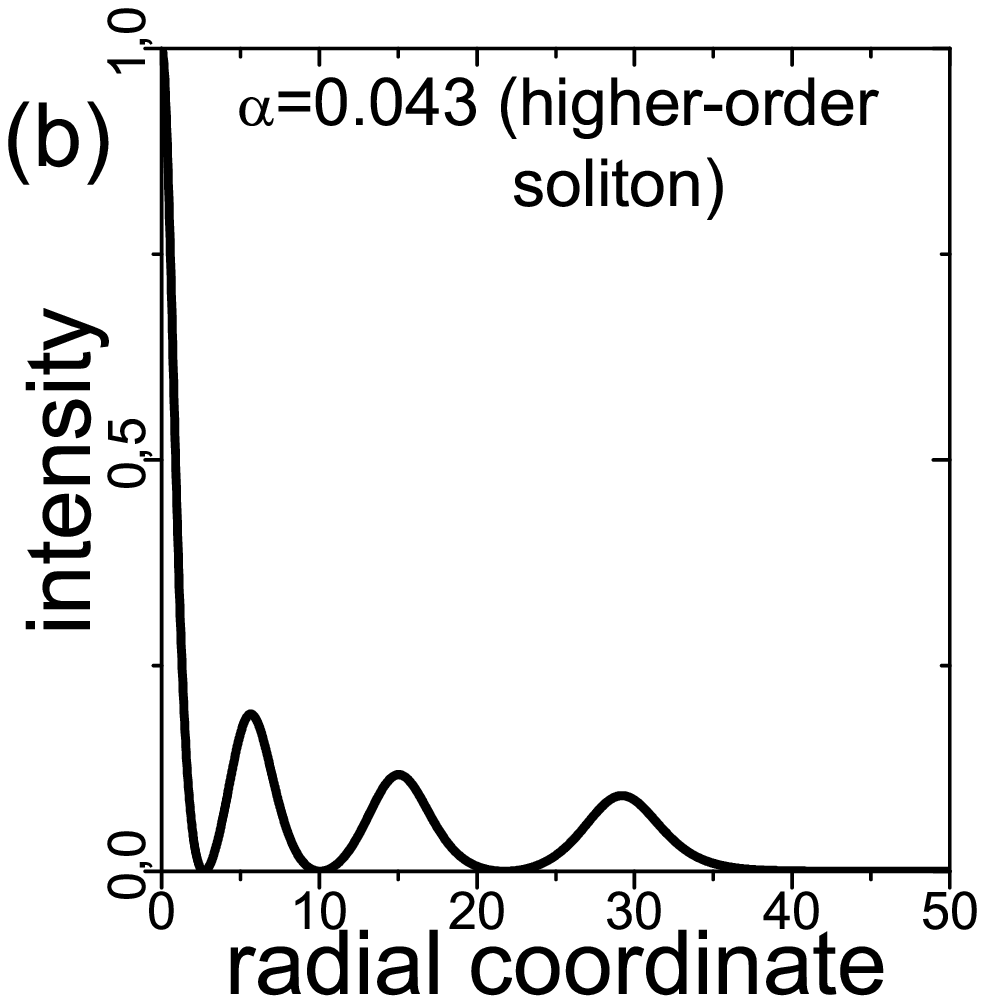}
\caption{\label{Fig1} Radial intensity profiles $|A|^2$ of (a) the Townes beam, (b) a higher-order spatial soliton. Intensity is normalized to the peak intensity $I_0$ and radial coordinate is $(k_0^2n_2I_0/n_0)^{1/2}r$, where $r=(x^2+y^2)^{1/2}$.}
\end{figure}

The solutions in the discrete spectrum are solitons (in a broad sense, since the rigorous notion of soliton requires also stability). For the NLSE in Eq. (\ref{NLSE}), the ground bounded spatial soliton in the discrete spectrum corresponds to $\alpha=0.205$, and is the well-known Townes beam \cite{CHIAO}. Higher-order bounded solutions correspond to isolated values of $\alpha$ approaching zero \cite{SOTO}. The radial profiles of the Townes beam and of some higher-order solitons are shown in Figs. \ref{Fig1}(a) and (b). Solitons are characterized by strong (exponential or faster) localization, by carrying finite energy, and by propagating at a subluminal phase velocity $v_p=\omega_0/(k_0+\delta)$, since $\delta>0$.

\begin{figure}
\centering
\includegraphics[width=4cm]{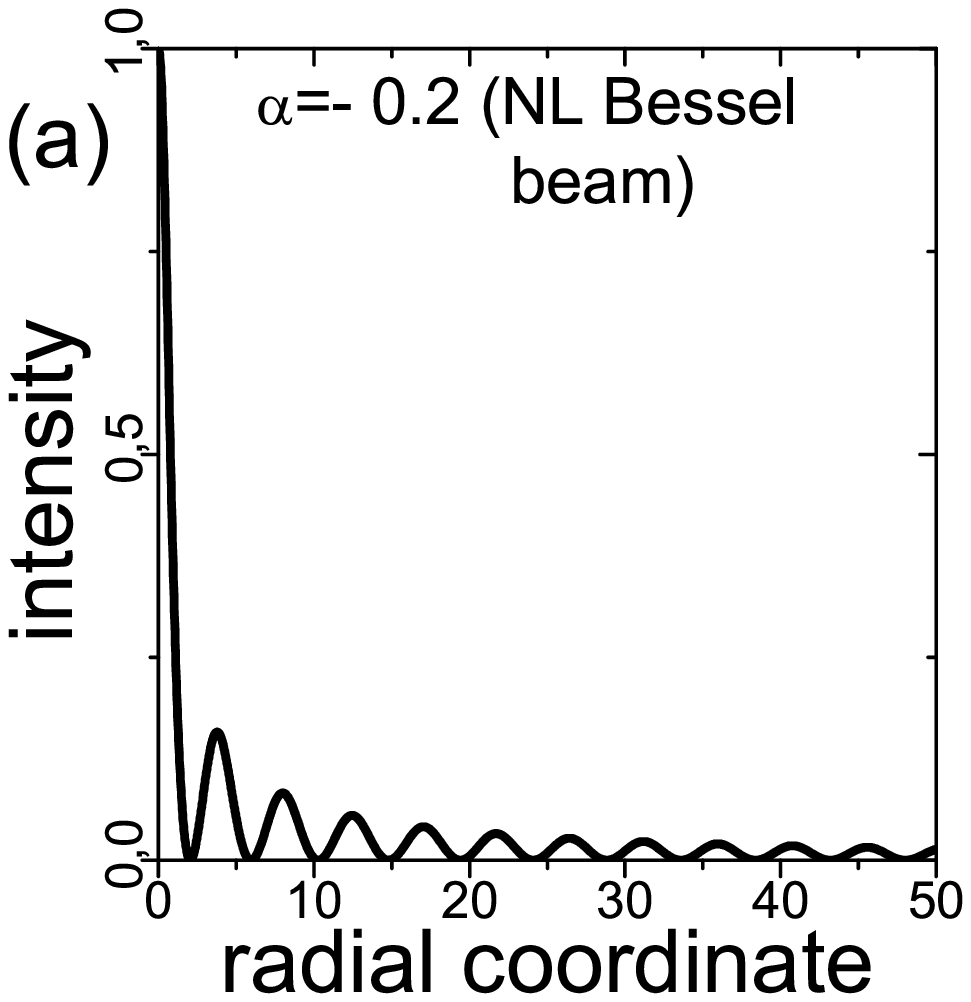}\includegraphics[width=4cm]{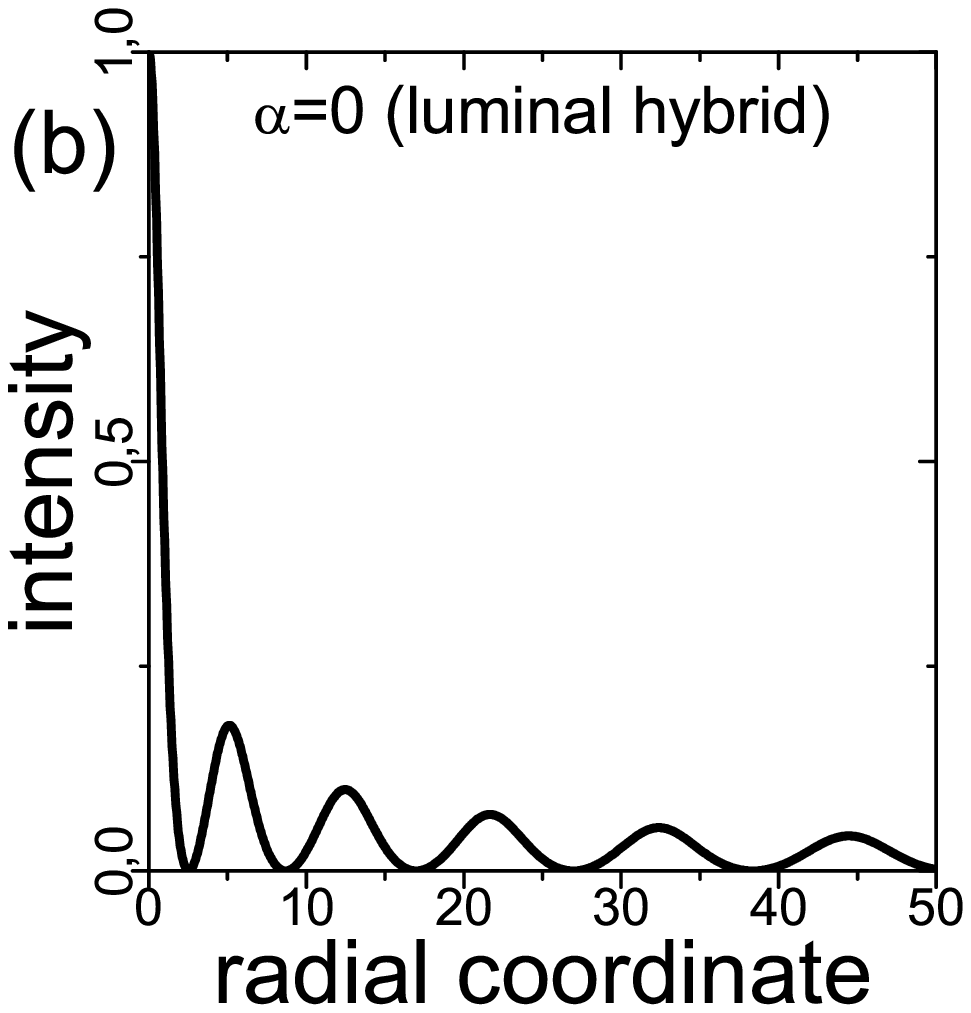}
            \caption{\label{Fig2} Radial intensity profiles $|A|^2$ of (a) a nonlinear Bessel beam, and (b) the luminal hybrid. Normalization of intensity and radial coordinate are as in Fig\ref{Fig1}.}
\end{figure}

The continuous spectrum with $\alpha<0$ has been less studied, probably because these stationary bounded solutions are weakly localized (decay algebraically) and carry infinite energy. Thus, only finite-energy versions are physically realizable. An example of their radial profiles is shown in Fig. \ref{Fig2}(a). In contrast to solitons, these wave packets are not fully nonlinear wave objects. They are called nonlinear Bessel beams \cite{STEPHANISEN} because they are generalizations to the nonlinear realm of the well-known non-diffracting Bessel beam \cite{DURNIN}. The intense center of nonlinear Bessel beams is affected by nonlinearities, but the periphery, containing most of the energy but widespread at low intensity behaves as the linear Bessel beam. Nonlinear Bessel beams have, as their linear counterparts, conical geometry, their cone angle $\theta$ being related to the axial wave vector shift by $\theta=\sqrt{-2\delta/k_0}$, and propagate at superluminal phase velocity.

More intringuing, and not reported up to now, is the bounded stationary solution of the NLSE with $\alpha=0$. It is an hybrid solitary-conical solution in the limit between the discrete and continuous spectra, whose profile [Fig. \ref{Fig2}(b)] is the limit those of solitons of infinitely high order and of nonlinear Bessel beams of cone angle approaching zero. This hybrid solution propagates without any axial wave vector shift, and therefore at luminal phase velocity, is weakly localization, and hence carries an unbounded amount of widespread energy. We are at present studying its propagation and stability properties.

The above spectral classification remains substantially unaltered for 3D self-localized wave packets, i.e., wave packets bounded not only in the two transversal directions, but also in time, but the variety of situations and waves increases, as well as their names. The case of a nonlinear medium with anomalous chromatic dispersion ($k_0^{\prime\prime}<0$) is particularly simple because dispersion plays the same role as diffraction ($\Delta$ contains three positive second derivatives, and hence the NLSE continues to be elliptic). The 3D, or spatiotemporal ground soliton, and higher-order solitons, similar to those in the discrete positive spectrum of the 2D case, are well-known many years ago. More recently, spatiotemporal nonlinear Bessel beams, called nonlinear O-waves, in the continuous negative spectrum have been reported, \cite{PORRASOWAVES1}, and the 3D, hybrid solution with $\alpha=0$ with luminal phase velocity is being currently studied. The case of normal dispersion is more complex due to the opposite signs in the dispersion and diffraction terms (the NLSE becomes hyperbolic), and a variety of solutions have been described. Among them, the generalizations of nonlinear Bessel beams are called nonlinear X waves, \cite{KOLESIK,TRAPANI3,TRAPANI1} and represent nonlinear counterparts of the linear X waves \cite{LINEARXWAVES,PORRASEX,PORRASMODES}. All these self-localized wave packets have been shown, or are supposed to be at work in experiments of spontaneous wave localization under different conditions \cite{KOLESIK,TRAPANI3,PORRASOWAVES2,TRAPANI1}.

Much more complex is the question of the stability of these self-localized waves. Analysis of stability under small perturbations is a relatively easy task for homogeneous waves \cite{LIOU}, but becomes barely tractable by analytical means for these inhomogeneous waves. Most conclusions must be drawn from approximate asymptotic methods \cite{ZAKHAROV,KUZNETSOV,PELINOVSKY,SKRYABIN,KIVSHAR} or numerical simulations \cite{SOTO,PORRASY}. 2D and 3D solitons in Kerr media are unstable because any perturbation makes them to self-focus catastrophically or diffract forever. Many mechanisms have been investigated and been shown to remove self-focusing instability, as saturation of the Kerr nonlinearity, \cite{SKARKA} higher-order chromatic dispersion \cite{FIBICH}, resulting in stable and self-localized propagation. The superluminal nonlinear Bessel beams in 2D, the O waves and X waves in 3D are supposed (as suggested by numerical simulations) to suffer also from a similar self-focusing instability, and the same is true for the luminal hybrid solutions in 2D and 3D.

\begin{figure}
\centering
\includegraphics[width=7cm]{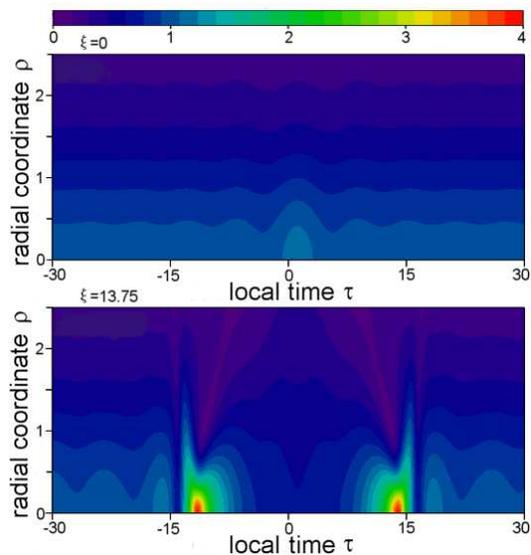}
\caption{\label{Fig3} Propagation of a temporally perturbed Townes beam. Top: Intensity of the initial perturbation in the form of a pulse on the top of the Townes beam. Bottom: Intensity of the propagated field with two X waves on the top of the partially depleted Townes beam. The radial coordinate is $(k_0^2n_2I_0/n_0)^{1/2}r$, where $r=(x^2+y^2)^{1/2}$, the local time is $(k_0n_2I_0/k_0^{\prime\prime}n_0)^{1/2} t'$, where $t'=t-k'_0z$, and the axial coordinate is $\xi= (k_0n_2I_0/n_0)z$. The intensity is normalized to that of the Townes beam.}
\end{figure}

Even if a self-localized wave has been stabilized \textit{in its own space}, it may suffer from instability \textit{in an extra dimension}, which is known as transversal instability \cite{KUZNETSOV,PELINOVSKY}. This point is particularly relevant for understanding experiments of self-focusing and filamentation with ultrashort pulses, as discussed below. For instance, in addition to its (removable) self-focusing instability, the Townes beam (and other spatial solitons) are temporally unstable \cite{PORRASY}. In a medium with normal dispersion, perturbation of a 2D soliton by a periodic temporal modulation results in its breaking into an infinite train of X waves \cite{PORRASU}. If the perturbation on the top of the 2D soliton is not periodic but localized in time, the perturbation develops into a couple of subluminal and superluminal X waves \cite{PORRASXYZ}. For illustration, Fig. \ref{Fig3}(top) shows a Townes beam (continuous in time and localized in the radial direction) with a weak perturbation about a given time. Figure \ref{Fig3}(bottom) shows that the weak perturbation grows on propagation at the expense of the Townes beam energy, and transforms gradually into two well-separated, subluminal and superluminal X waves. This situation resembles the phenomenon of pulse splitting in light filaments in media with normal dispersion, and points out a relevant role of conical waves in filamentation. \cite{KOLESIK,TRAPANI3}

\section{Resistance to dissipation and self-reconstruction property}

The crucial question is which of these self-localized waves, and which stabilizing mechanisms, are involved in experiments of filamentation, or could be practiced in future experiments for the generation of light-bullets. Most of experiments in this respect consist on letting a standard Gaussian wave packet self-focus in a nonlinear medium \cite{SILBERBERG1,SILBERBERG2}. Depending on its optical properties and on the input conditions (wavelength, duration, power \dots ) collapse is halted and the compressed wave packet stabilized by one or several optical effects other than Kerr nonlinearity, as normal group velocity dispersion, saturation of the Kerr nonlinearity, multiphoton ionization,plasma defocusing, etc. \cite{SKARKA,FIBICH,HENZ,BRAUN}

An unexpected player in this problem is dissipation of energy. For the extremely high intensities involved in filamentation experiments, nonlinear mechanisms of dissipation of energy are turned \cite{SILBERBERG2,HENZ,BRAUN,MOLL,BERGE}. Several photons can be absorbed in a multiphoton process so as to ionize the atoms, and to create an electron plasma that in turn modifies the properties of the medium. These nonlinear losses are known to halt collapse, but were not suspected some years ago to be compatible with stationary and stable wave packet propagation \cite{PORRASUBB}.

The NLSE
\begin{equation}\label{NLSENLL}
\partial_z A=\frac{i}{2k_0}\Delta A  +\frac{ik_0n_2}{n_0}|A|^2A-\frac{\beta^{(K)}}{2}|A|^{2K-2}A\,
\end{equation}
with nonlinear absorption ($\beta^{(K)}>0$ is the multiphoton absorption coefficient, and $K$ is the order of the process), no longer have a discrete spectrum of bounded stationary solutions. i.e., does not support solitary wave packets. However the continuous spectrum $\alpha<0$ of conical waves, and the limiting hybrid solution with $\alpha=0$, continue to exist. These solutions have the capability of propagating in a nonlinear lossy medium without any distortion of attenuation.

\begin{figure}
\centering
\includegraphics[width=8.5cm]{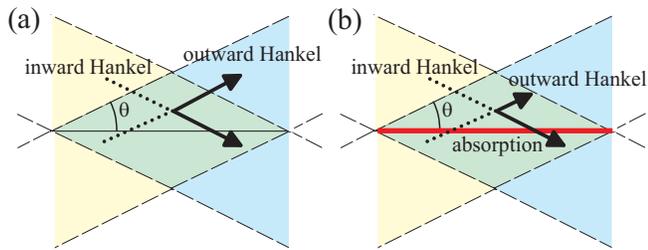}
\caption{\label{Fig4} Understanding the replenishment mechanism of nonlinear unbalanced Bessel beams that makes them resistant to nonlinear losses. (a) Linear Bessel beam. (b) Nonlinear unbalanced Bessel beam.}
\end{figure}

\begin{figure}
\centering
\includegraphics[width=6cm]{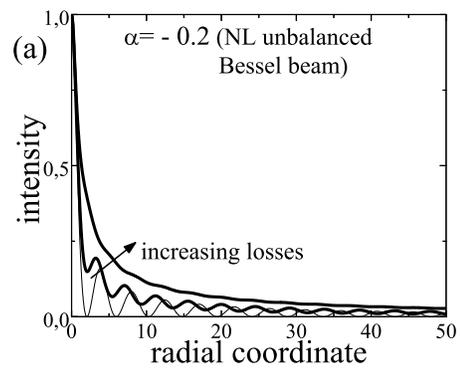}
\caption{\label{Fig5} Radial intensity profiles $|A|^2$ of nonlinear unbalanced Bessel beams of given cone angle and increasing losses. The thin curve corresponds to the nonlinear Bessel beam without losses. Intensity and radial coordinates are normalized as in Fig. \ref{Fig1}.}
\end{figure}

The simplest member of these solutions of the NLSE (\ref{NLSENLL}) is the so-called nonlinear unbalanced Bessel beam (NL-UBB) in 2D \cite{PORRASUBB}. Its resistance to nonlinear losses can be understood from Fig. \ref{Fig4}. A linear Bessel beam is well-known to consists of the superposition of two more elemental conical beams, called inward and outward Hankel beams. The inward Hankel beam is endowed of a conical energy flux toward the propagation axis $z$, and the outward Hankel beam from the propagation axis. In the linear Bessel beam (and in the nonlinear Bessel beam far from its center) propagating in transparent media, the two Hankel beams have equal amplitudes [Fig. \ref{Fig4}(a)], which results in a null transversal energy flux. In the nonlinear unbalanced Bessel beam in a medium with nonlinear losses, the energy is continuously being absorbed in its intense peak around the propagation axis. The amplitude of the outward Hankel beam is then decreased compared to that of the inward Hankel beam [Fig. \ref{Fig4}(b)]. This results in a net energy flux toward the central peak that refuels the energy in it. Unbalancing of the two Hankel components is manifested as a reduction of the contrast of the radial oscillations in the radial intensity profile compared to those of Bessel beams (Fig. \ref{Fig5}). The higher the losses (e.g., the peak intensity), the more pronounced the unbalancing needed for replenishment and the lower the contrast.
Clearly, stationarity with nonlinear losses relies on the property of conical waves of having an infinite reservoir of widespread energy in its tails. Physically realizable, nonlinear unbalanced Bessel beams with finite energy (e.g., truncated) have been seen to resist also nonlinear losses, but only up to the distance at which their finite-energy reservoir is consumed, a distance that can exceed by hundred of times the diffraction length corresponding to the width of its central peak \cite{PORRASUBB}.

Similar bounded and stationary solutions of the NLSE (\ref{NLSENLL}) in three dimensions in media with anomalous dispersion have been described \cite{PORRASOWAVES1,PORRASOWAVES2}. They generalize the linear and nonlinear O-waves in transparent media to the lossy case, and have been named nonlinear unbalanced O-waves. Similar solutions are supposed to exist in media with normal dispersion, but have not been reported yet.

These results have led to the general belief that resistance to losses is an privative property of conical waves due to its peculiar conical geometry. However, the luminal hybrids with zero cone angle ($\alpha=0$) in 2D and 3D are also solutions of the NLSE (\ref{NLSENLL}) with nonlinear losses that represent localized and undistorted propagation in absorbing media. In this case, replenishment is supplied by an inward energy flux created by a permanent, but frozen, state of self-focusing \cite{PORRASAPB}.

\begin{figure}[!b]
\centering
\includegraphics[width=4cm]{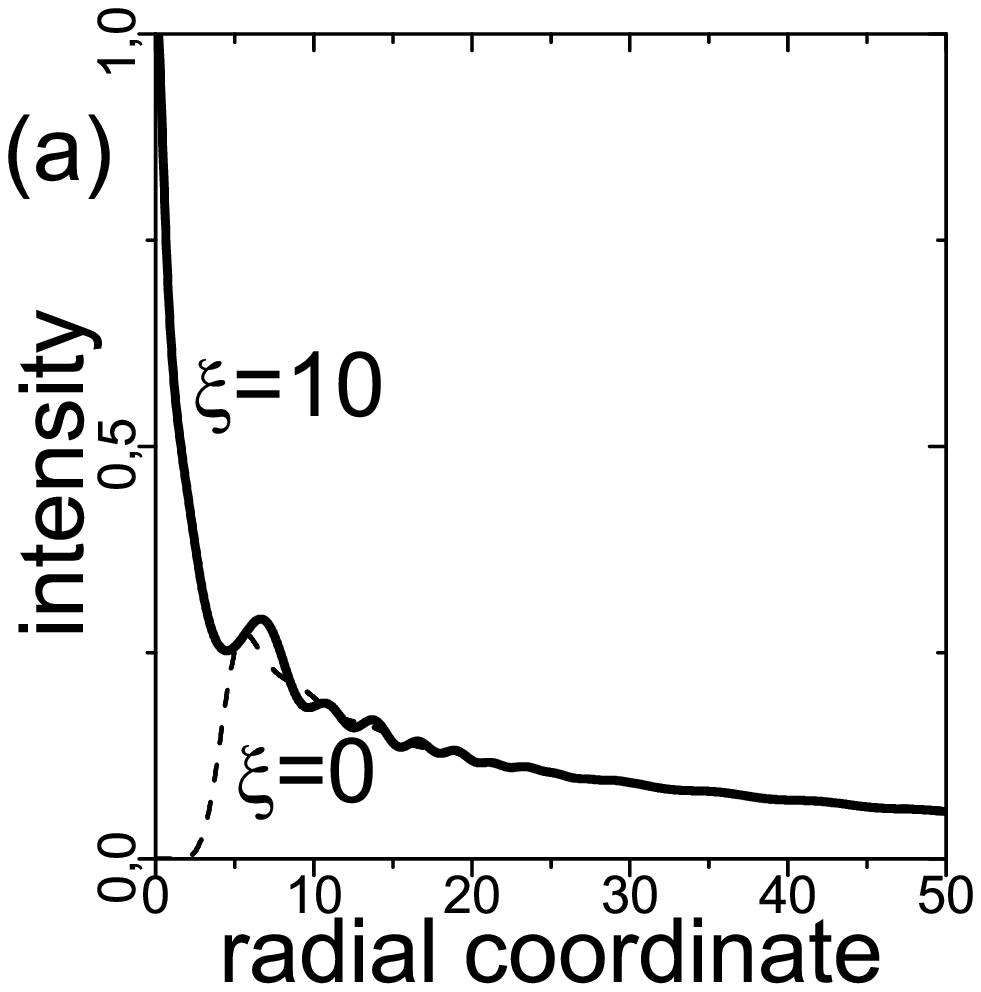}
\includegraphics[width=4cm]{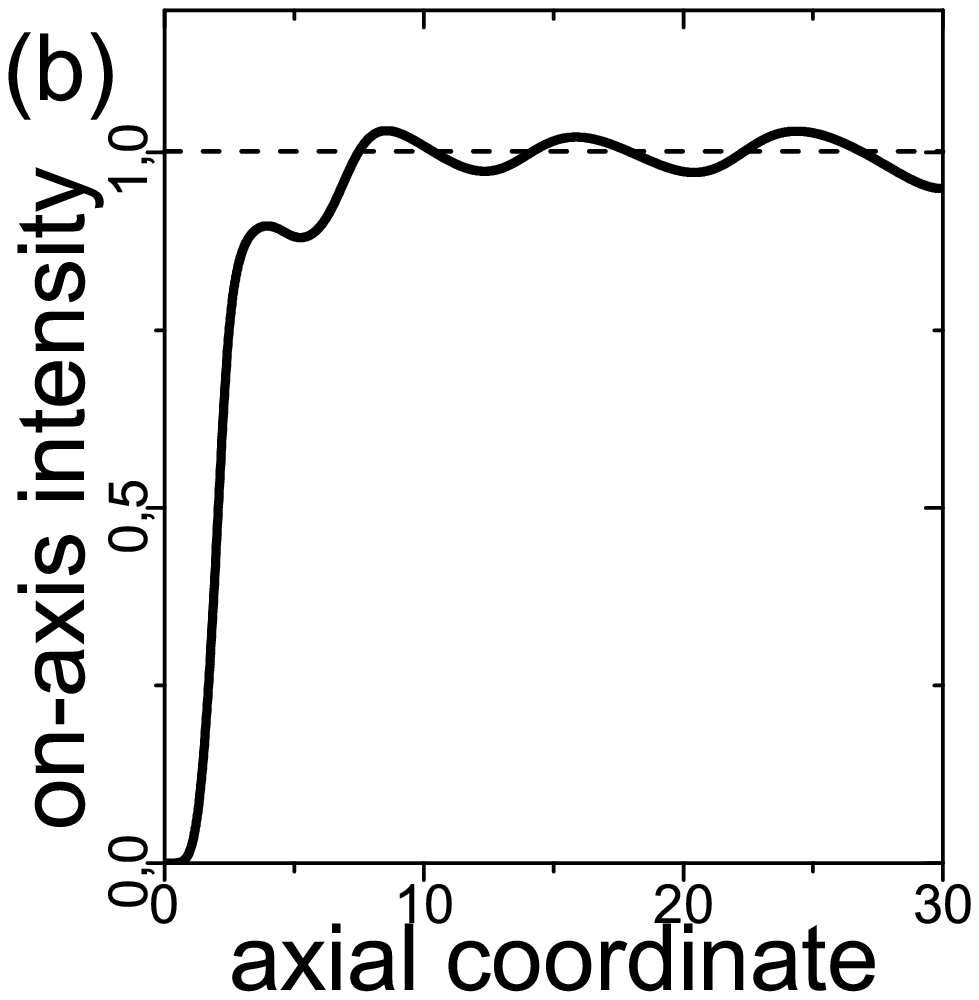}
\caption{\label{Fig6} Radial intensity profiles $|A|^2$ of a 2D luminal hybrid whose central peak is removed (dashed curve) and of the same beam after some propagation distance (solid curve). (b) Peak intensity as a function of the propagation distance.
Intensity, radial and axial coordinates are normalized as in the preceding figures.}
\end{figure}

Closely connected to resistance to losses is the self-reconstruction property of superluminal conical and luminal hybrids after obstacles. This property was first described for linear Bessel beams \cite{BOUCHAL}, but have been seen to hold equally for nonlinear Bessel beams \cite{DUBIETIS}, and more surprisingly, for their limit of zero cone angle, i.e., for the luminal hybrids \cite{PORRASAPB}. For instance, the dashed curve in Figure \ref{Fig6}(a) is the radial profile of a 2D luminal hybrid whose central peak of intensity is absorbed by an obstacle, and the solid curve represents the same beam with the self-reconstructed peak after some propagation distance. The peak intensity along the propagation axis is seen in Fig. \ref{Fig6}(b) to recover and stabilized into its unblocked value, as if there had been no obstacle.

\section{Spontaneous localization in light filaments}

Many experiments of spontaneous wave localization in homogeneous media involve careful self-focusing in gases, liquids or solids, of ultrashort pulses (of few hundreds of femtoseconds) with power well above the critical power for self-focusing. Arrest of collapse leads to the formation of one or several self-channeled wave packets, or light filaments, that can propagate without apparent change for very long distances. A detailed analysis of the light filament dynamics requires much more accurate equations than the simple NLSE  (\ref{NLSE}) in order to include higher-order linear and nonlinear effects, as Kerr saturation and shock terms, nonlinear absorption, higher-order dispersion, and coupling to the electron plasma generated by multiphoton or avalanche ionization (see, for instance \cite{COUAIRON}).

There is still a debate about the self-trapping mechanism in these light filaments, since they exhibit, behind an apparent stationarity, a rich  dynamics in space and time. The formation of a light filament involves spectral broadening over the whole visible spectrum and beyond. In media with normal dispersion, the collapsing pulse splits into two pulses that propagate with different group velocities along the filament path, and that may recombine to self-focus once more, or many times. Further, light filaments emit a kind of Cerenkov radiation, called conical emission, observable on a screen intercepting the filament as colored rings. \cite{COUAIRON}

The most widely accepted explanation of the stable self-localization of light filaments relies on the soliton concept. The electron plasma generated around the collapse region lowers the refraction index, creating a self-defocusing effect that balances stably Kerr self-focusing \cite{COUAIRON,HENZ,BRAUN}. However, this explanation does not incorporate the dynamic aspects of the filament. Also, recent experiments showed that light filaments self-reconstruct after being blocked by obstacles (e.g., by water drops in air, or stoppers placed by purpose in water) \cite{DUBIETIS,DUBIETIS2,KOLESIK2}, a property that is hardly reconciled with the idea of a soliton.

These limitations have led to alternate explanations in which plasma defocusing is not the key player in self-localization. The apparent stationarity is reconciled with the dynamic aspects in the theory of dynamical nonlinear X waves \cite{KOLESIK}. The precise characterization of the conical emision in experiments of filamentation in media with normal dispersion revealed that conical emision is the result of the formation of two, superluminal and subluminal X waves that co-propagate with the split off pulses \cite{TRAPANI3}. Each X wave is formed as a result of the parametric amplification, in a four-wave mixing proccess driven by the Kerr nonlinearity, of idler and signal spatiotemporal frequencies by each pump or split off pulse. The spatiotemporal frequencies favored by the phase matching conditions of this process are just those in the spectrum of the X waves \cite{TRAPANI3}.

\begin{figure}[!tbp]
\centering
\includegraphics[width=7cm]{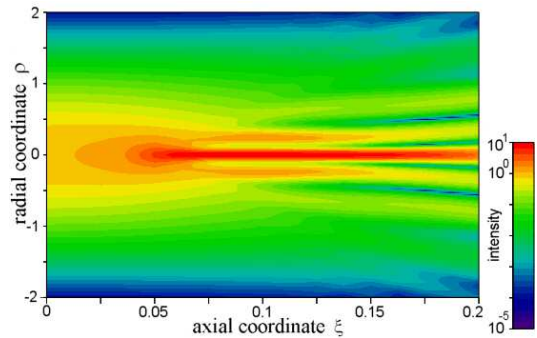}
\caption{\label{Fig7} Self-focusing of a Gaussian wave packet in a Kerr medium with nonlinear losses, and filamentation into a finite-energy version of the luminal hybrid resistant to losses. The filament survives until the energy reservoir refilling the central peak is consumed.}
\end{figure}

This theory still does not explains the origin of pulse splitting. A theory that incorporates this phenomenon, and that reconciles the interpretations  in terms of solitons and of conical waves, is that spectral broadening, pulse splitting and conical emission as two subluminal and superluminal X waves, are manifestations of the temporal instability of spatial solitons with normal dispersion \cite{PORRASY,PORRASXYZ,PORRASU}. Arrest of collapse implies the stabilization \textit{ in space} of the self-focusing pulse into a kind of spatial soliton with an approximate Townes radial profile \cite{MOLL2}. However, the development of instability \textit{in time} results, as discussed in Sec. 2, in spectral broadening, and in splitting of the soliton into the two subluminal and superluminal X waves that are observed as conical emission \cite{PORRASXYZ}.

Filamentation with anomalous dispersion has deserved less attention. Diffraction and dispersion play here a similar role, which prevents pulse splitting to occur, and facilitates symmetric spatiotemporal self-focusing into light bullets in three dimensions. \cite{SILBERBERG1,MOLL} Filament paths have indeed been shown to be much longer compared to case with normal dispersion. Though the higher intensities involved makes dissipation by multiphoton ionization to play a prominent role in the arrest of collapse, plasma defocusing continues to be considered  responsible for stable self-guiding of the presumed soliton-like filament, since nonlinear losses are not considered be compatible with stationary propagation. We have shown recently \cite{PORRASEXX} that spatiotemporal self-focusing halted by nonlinear losses leads to the formation of a nonlinear unbalanced O wave resistant to losses with a cone angle approaching zero, i.e., to the formation of the 3D luminal hybrid in which a permanent state of self-focusing refuels the energy continuously lost during propagation, and that the formation of this luminal hybrid explains many of the observed features of light filaments with anomalous dispersion \cite{MOLL,BERGE}. The formation of a light filament does not require the defocusing effect of plasma. Obviously, only a finite-energy version of the 3D luminal hybrid is formed, since the energy of the input self-focusing pulse is finite. The filament decays when the finite-energy reservoir of the luminal hybrid is consumed \cite{PORRASEXX}. Figure \ref{Fig7} illustrates the spontaneous formation of the luminal hybrid in the self-focusing of an input Gaussian wave packet halted by nonlinear losses, as obtained by direct numerical integration of the NLSE (\ref{NLSENLL}). \cite{PORRASSP}

\section{Conclusion and acknowledgements}

I conclude by stressing that this research has been conducted in close collaboration with other groups and researchers, as seen in the reference list. I am particularly indebted to Alberto Parola for acting as my teacher in nonlinear wave phenomena and in numerical simulations.

\end{document}